\documentclass[12pt,a4paper]{article}
\usepackage{jheppub}
\usepackage{amssymb,amsmath,amsfonts}
\usepackage{epsfig}

\def\clock{{\count0=\time
           \divide\count0 60
           \ifnum\count0<10 0\fi\the\count0
           \multiply\count0 -60 \advance\count0 \time
           :\ifnum\count0<10 0\fi \the\count0
         }}
\newcommand{\timestamp}{{\small\vbox{\hbox{\tt\jobname.tex}
\hbox{\the\day/\the\month/\the\year, \clock}}}}

\newcommand{\CL}{\mathcal{L}}

\newcommand{\nn}{\nonumber}

\newcommand{\ads}{\mbox{AdS}}

\makeatletter
\def\mr@ignsp#1 {\ifx\:#1\@empty\else #1\expandafter\mr@ignsp\fi}%
\newcommand{\multiref}[1]{\begingroup
\xdef\mr@no@sparg{\expandafter\mr@ignsp#1 \: }%
\def\mr@comma{}%
\@for\mr@refs:=\mr@no@sparg\do{\mr@comma\def\mr@comma{,}\ref{\mr@refs}}%
\endgroup}
\makeatother

\newcommand{\hypref}[2]{\ifx\href\asklfhas #2\else\href{#1}{#2}\fi}

\renewcommand{\eqref}[1]{(\multiref{#1})}

\def\[{\begin{equation}}
\def\]{\end{equation}}
\newcommand{\be}{\begin{eqnarray}}
\newcommand{\ee}{\end{eqnarray}}

\def\th{\theta}

\newcommand{\p}{\partial}
\newcommand{\half}{\frac{1}{2}}
\newcommand{\quarter}{\frac{1}{4}}

\newcommand{\alg}[1]{\mathfrak{#1}}
\newcommand{\grp}[1]{\mathrm{#1}}

\newcommand{\grSU}{\grp{SU}}
\newcommand{\grSO}{\grp{SO}}

\newcommand{\algSU}{\alg{su}}

\newcommand{\sphere}{\mbox{S}}
\newcommand{\cft}{\mbox{CFT}}
\newcommand{\lmtilde}{\tilde{\lambda}}
\newcommand{\smtilde}{\tilde{\sigma}}

\def\tn{\textnormal}

\def\bea{\begin{eqnarray}}
\def\eea{\end{eqnarray}}

\usepackage{color}

\title{ The dual string $\sigma$-model of the $\grSU_q(3)$ sector }
\author[a]{Valentina Giangreco M. Puletti}
\author[b]{and Teresia M\aa{}nsson}

\affiliation[a]{NORDITA\\
Roslagstullsbacken 23,
SE-106 91 Stockholm,
Sweden}

\affiliation[b]{KTH (Royal Institute of Technology) \\
Roslagstullsbacken 21,
SE-106 91 Stockholm,
Sweden}

\emailAdd{valentina@nordita.org}
\emailAdd{teresiam@kth.se}

\abstract{
In four-dimensional $\mathcal N=4$ super Yang-Mills (SYM) the $\grSU(3)$ sub-sector spanned by purely holomorphic fields is isomorphic to the corresponding mixed one spanned by both holomorphic and antiholomorphic fields. 
This is no longer the case when one considers the marginally deformed $\mathcal{N}=4$ SYM. The mixed $\grSU(3)$ sector marginally deformed by a complex parameter $\beta$, {\it i.e.} $\grSU_q(3)$ with $q=e^{2 i\pi\beta}$, has been shown to be integrable at one-loop~\cite{Mansson:2007sh}, while it is not the case for the corresponding purely holomorphic one.
Moreover, the marginally deformed $\mathcal N=4$ SYM also has a gravity dual constructed by Lunin and Maldacena in~\cite{Lunin:2005jy}. However, the mixed $\grSU_q(3)$ sector has not been studied from the supergravity point of view. Hence in this note, for the case of purely imaginary marginal $\beta$-deformations, we compute the superstring $\grSU_q(3)$ $\sigma$-model in the fast spinning string limit and show that, for rational spinning strings, it reproduces the energy computed via Bethe equations. 
}

\rightline{\vbox{\tiny\hbox{NORDITA-2011-44} }}
\begin{document}
\maketitle
\flushbottom

\setcounter{page}{1}


\section{Introduction}
\label{sec:Intro}

The $\ads_5/\cft_4$ correspondence conjectures a duality between $\mathcal N=4$ $\grSU(N)$ super Yang-Mills (SYM) in four
dimensions and type IIB superstring theory on $\ads_5 \times \sphere^5$~\cite{Maldacena:1997re, Gubser:2002tv, Witten:1998qj}.
Since the duality relates weak and strong regimes of the two different theories, it is not provable by pertubative methods alone.
The discovery of an infinite number of conserved charges ({\it i.e.} integrability) on both sides of the $\ads_5/\cft_4$
correspondence ~\cite{Minahan:2002ve,Beisert:2003yb,Beisert:2003tq, Kazakov:2004qf,Arutyunov:2004vx,Beisert:2006ez}%
 ~\footnote{Cf. \cite{Beisert:2010jr} for a comprehensive and recent list of references on the topics.}
has led to much progress both regarding attempts to verify the conjecture and applications.
The existence of such symmetries
has allowed us to understand various connections between spin chains, AdS superstring theories and gauge theories. Nowadays
we therefore have strong evidence that $\mathcal N=4$ SYM is solvable at least in the large $N$ regime. Integrability is an extremely powerful tool for studying and applying of the duality. It would be desirable if it could also be
applied to further the progress on less supersymmetric gauge theories,%
\footnote{We recall that $\mathcal N=4$ SYM in four dimensions is a maximally supersymmetric theory (32 supercharges). The
$\ads_5\times \sphere^5$ background of the dual type IIB superstring theory also preserves 32 supersymmetries.}
having in mind QCD as a long-term prospect.%
\footnote{
For interesting works in this direction, we refer the reader to \cite{Ferretti:2004ba,Beisert:2004fv} and to the recent review \cite{Korchemsky:2010kj} in the context of QCD, to \cite{DiVecchia:2004jw,DiVecchia:2004xb} for  works related to $\mathcal N=2$ SYM, and to \cite{Gadde:2009dj,Gadde:2010zi,Gadde:2010ku, Pomoni:2011jj,Liendo:2011xb,Poland:2011kg} for works related to $\mathcal N=2$ superconformal QCD. 
}

In this context, one of the most well-studied examples of less supersymmetric theories are the Leigh-Strassler deformed
gauge theories~\cite{Leigh:1995ep}. These are marginal deformations of $\mathcal{N}=4$ SYM. In the $\mathcal N=1$ superspace
formulation, the corresponding most general deformed superpotential is given by
\be
\label{superpotential}
W=\kappa\, \mathrm{Tr}\left (\, \Phi_1\, [\Phi_2\,,\,\Phi_3]_q + {h\over 3} \left( \Phi_1^3+\Phi_2^3 +\Phi_3^3\right)\right)\,,
\ee
where the $q$-commutator is defined as
\be
[A,B]_q\,=\,A\,B-q\,B\, A\,,
\ee
and $q$, $\kappa$ are in general complex parameters. Here, $\Phi_1\,, \Phi_2$ and $\Phi_3$ are the three complex scalars of
SYM spanning the $\grSU(4)$ sub-sector. In the case we are interested in, {\it i.e.} $h=0$ and $N\rightarrow \infty$, $\kappa$
is%
\footnote{This is the condition for finiteness of the Leigh-Strassler deformed $\mathcal{N}=4$ SYM to second order in
perturbation theory in the coupling constant $g$ \cite{Jones:1984cx,Parkes:1984dh}, but also the condition for the
$\grSU(3)$ sub-sector made up by two holomorphic and one anti-holomorphic scalar field to be integrable at one loop
\cite{Mansson:2007sh}.}
\be\label{def_kappa}
\kappa\bar{\kappa}=\frac{2g^2}{1+q\bar{q}}\,,
\ee
where $g$ is the SYM coupling constant. Usually, $q$ is defined in terms of a complex parameter $\beta$ as 
\be
q=e^{2i\pi\beta}\,,
\ee
thereby the often used terms $\beta$-deformations and $q$-deformations.%
\footnote{For a very recent work on Leigh-Strassler deformations with non-zero $h$, we refer the reader to~\cite{Minahan:2011dd} and references therein.}%
 

The marginal deformations of $\mathcal N=4$ SYM preserve the conformal symmetry of the undeformed theory, but
they affect the number of supersymmetries. As already mentioned, the study of the deformed $\ads_5/\cft_4$
correspondence offers an important testing ground for the gauge/gravity duality with a reduced number of
supersymmetries. From the very beginning there have been considerable efforts to understand the gravitational
dual of the marginal deformed SYM. In \cite{Lunin:2005jy}, Lunin and Maldacena constructed such a dual background
for any complex $q$, for $h=0$ and for $\kappa$ obeying \eqref{def_kappa}. For the specific case of real $\beta$,
the gravitational background is obtained by a series of transformations involving T-duality,
a shift of an $\sphere^5$ angle and again a T-duality (TsT transformations). When $\beta$ is complex, 
it is necessary to perform additional S-duality transformations.

The integrable structures of the planar $\mathcal N=4$ SYM and of its gravitational
dual survive to some extent also in the corresponding marginally deformed
theories. This is a highly non-trivial statement since the deformations break some (or even all) supersymmetries,
and at the quantum level the original theory and the deformed one are very different. Thus, there is no guarantee that
the emerging solvability of $\mathcal N=4$ SYM should be inherited by the deformed SYM. 

In this direction, a first step was made in the milestone paper~\cite{Roiban:2003dw}, where it was shown that, at one-loop order, the $\beta$-deformed
$\grSU(2)$ theory is integrable for general complex $\beta$.
In particular, it was shown that the one-loop dilatation operator in the holomorphic two-field sub-sector
corresponds to the integrable XXZ-spin chain.
For the real $\beta$-deformed $\grSU(2)$ sub-sector, an important comparison between gauge and string theory was
made in \cite{Frolov:2005ty}, where it was shown that the fast spinning string $\sigma$-model matched the coherent
state action of the gauge theory side. 

Outside the $\grSU(2)$ sub-sector, the paper~\cite{Roiban:2003dw} also discussed the possibility
that integrability might survive for a deformed $\grSU(3)$ sector. In \cite{Berenstein:2004ys}, 
it was understood that integrability is only preserved for special values of complex $q$ ($\beta \in \mathbb{R}$)
when one considers the deformation of the holomorphic $\grSU(3)$ sector. 
A further progress was made in \cite{Mansson:2007sh}: There it has been shown that the deformed $\grSU(3)$ sub-sector spanned by an anti-holomorphic and two holomorphic fields (or vice versa) is integrable for any value of $\beta$. This is the sub-sector we consider in this work, but we restrict ourselves to the case of purely imaginary $\beta$.


A generalization of the examples discussed above are the so-called $\gamma_i$ deformations of SYM~\cite{Frolov:2005dj}:
Different TsT transformations are performed along the three $\grp{U}(1)\subset \grSU(4)$ directions of the 5-sphere $\sphere^5$.
This breaks all the supersymmetries, but nevertheless the authors of~\cite{Frolov:2005iq} proved that the theory is still
integrable by matching the fast spinning string $\sigma$-model action constructed in this background with the coherent state spin
chain Hamiltonian. They also computed the corresponding Bethe equations and for special string configurations they showed that the two spectra 
obtained in this way agree completely. All the results contained in \cite{Frolov:2005iq}
are valid only when all the three parameters $\gamma_i$ are real. 
Furthermore, the same paper \cite{Frolov:2005iq} investigated the extension of the fast spinning string action to the
complex $\beta$ case. However, no comparison could be made there between gauge and string theory, 
since no Bethe equations were formulated for the $\grSU(3)$ sector for complex $\beta$.%
\footnote{The Bethe equations formulated in~\cite{Beisert:2005if} for the real $\beta$ case, when extended to complex
values of $\beta$, do not produce any real eigenvalues.}
For more details, we refer the reader to the recent review~\cite{Zoubos:2010kh} and references therein.


Along this line of research, the aim of this work is to continue the study of integrability beyond the $\grSU(2)$
sub-sector in the marginal deformed $\ads_5/\cft_4$ duality. In particular, we restrict to imaginary $\beta$ and to the 
$\grSU_q(3)$ sub-sector spanned by one anti-holomorphic and two holomorphic fields. This will be called the
mixed sector in the rest of the paper. We derive the string $\sigma$-model describing the fast rotating string in
this mixed $\grSU_q(3)$ sector of the Leigh-Strassler deformations. Following the same approach%
\footnote{For the undeformed $\ads_5/\cft_4$ case, fundamental works in this direction are
\cite{Kruczenski:2003gt, Kruczenski:2004kw}.}
as in \cite{Frolov:2005iq}, this action can be used to calculate the energy corresponding to the rational spinning
string solution. Eventually, this is compared to and agrees with the predictions we obtain from the Bethe equations,
which were determined in \cite{Mansson:2007sh} for the mixed and deformed $\grSU_q(3)$ sector. 

In order to directly compare string and gauge theory results, one should re-sum all the planar diagrams or, equivalently, consider the full planar S-matrix, which could have a non-trivial phase factor, {\it i.e.} the so-called dressing phase \cite{Beisert:2006ez}.
However, at the leading order in $\lambda'$ in the undeformed $\ads_5/\cft_4$ and in the marginally deformed $\ads_5/\cft_4$ for real values of the parameter $\beta$, the dressing factor does not affect the anomalous dimension of infinitely long single-trace operators as well as the spectrum of infinitely long strings. 
We do expect that a similar situation is true also in the marginally deformed $\ads_5/\cft_4$ for imaginary values of $\beta$, namely for the case studied in this paper. For this reason, we directly compare strongly and weakly coupled regimes and we adopt the same semi-classical approach as in the early days of integrability. Indeed, the results we find, confirms our intuition in a non trivial way.

In $\mathcal{N}=4$ SYM there is no difference between the pure holomorphic and the mixed $\grSU(3)$
sector due to the $\grSO(6)$ invariance. However, when one marginally deforms $\grSO(6)$, the holomorphic
and the mixed sectors are no longer identical. This difference allows us to compare (for the mixed
$\grSU_q(3)$ sector) the energy obtained from the fast spinning string with the one-loop anomalous dimension of
the corresponding dual gauge theory operators. In fact, adding a non-holomorphic
field to the theory leads to a sub-sector which is integrable and closed up to one-loop, namely the $\algSU_q(3)$ sector~\cite{Mansson:2007sh}.

%
%
%
%
In~\cite{Avan:2010xd}, the continuum limit for a large class of spin chain Hamiltonians has beeen considered,
including the one describing the one-loop dilation operator investigated here. However, the approach of~\cite{Avan:2010xd} lacks a term which
is quadratic in the small deformation parameter. One can understand that
such a term should be present from the Bethe equations, as we explain in Section \ref{sec:BE}.
 We refer to this missing term as a
``cubic'' term, since it is the product of the three components of an $\grSU(3)$ coherent state up to phases
({\it i.e.} the three components of the magnetization vector in the spin chain language).
 One can check that the long wave length 
Hamiltonian~\cite{Avan:2010xd} agrees with the spin chain model obtained via the $\grSU(3)$ coherent states
(we review this method in appendix \ref{sec:coherent_sigma_mod}).
 Since the spin chain Hamiltonian is not $\grSU(3)$ invariant, it is not unexpected that this method will not give the complete answer.%
\footnote{In the purely holomorphic deformed $\grSU(3)$ sector a similar issue was already pointed out in \cite{Frolov:2005iq}. }
However,
what is a surprise to us is that all 
the terms quadratic
in the deformation parameter, except for the cubic one ($\rho_1^2\rho_2^2\rho_3^2$), are present. 
In this example, supergravity leads us to the correct long wave length limit of the spin chain Hamiltonian.

\vskip 1 cm
\subsubsection*{Outline}

In section \ref{sec:sigma_model} we develop a general treatment of the fast spinning string action and we apply the formula to the Lunin-Maldacena background with the special choice
for the parameters corresponding to the mixed deformed $\grSU_q(3)$ sector. Further, we compare the fast spinning string
$\sigma$-model in the $\grSU(3)$ deformed mixed sector with the one corresponding to the deformed holomorphic one. We also discuss the three spin rational solution.

In section \ref{sec:BE} we find a solution to the Bethe equations corresponding to the rational spinning string solution. We show
that such a solution agrees with the one obtained on the string theory side in section \ref{sec:general_fast_string}.

The first appendix \ref{sec:geom_setup} summarizes the notation and the geometrical set-up corresponding to the Lunin-Maldacena background. The second and third appendices, \ref{sec:explicit_derivation} and \ref{subsec:Solving the Bethe equations}, are devoted to collect the explicit computations which have been omitted in the main text.
Finally, in appendix \ref{sec:coherent_sigma_mod} we consider the coherent sigma model which one
obtains using the naive $\grSU(3)$ coherent states, and we see that it is the same up to a cubic
term with the action obtained from the string theory side in section \ref{sec:sigma_model}.


\section{The $\grSU_q(3)$ string $\sigma$-model}     
\label{sec:sigma_model}

\subsection{Preliminaries}
\label{sec:general_fast_string} 
 
 In this section we derive the effective action for a string, which is entirely moving in the compact deformed sector $\sphere^5$ with a total angular momentum $J$ and which is point-like in $\ads_5$. In the limit we are considering, the string is extended and $J$ is very large, namely the string is fast spinning in the deformed $\sphere^5$.
Firstly, we need to identify and gauge away the fast collective coordinate associated to the angular momentum $J$ and then, expand the slow transverse coordinates in powers of an effective parameter $\lmtilde\equiv \lambda/J^2$. Notice that both $\lambda$ and $J$ are very large, but their ratio is fixed, and in particular small. Thus, the resulting low-energy $\sigma$-model is only expressed in terms of the transverse string coordinates, which are usually interpreted as the components of a classical spin vector belonging to $\grSU(3)$ in this case (\emph{e.g.} the classical magnetization in the Landau-Lifshitz model). For more details cf. \cite{Tseytlin:2004xa, Stefanski:2004cw}. 
  
In order to derive the $\sigma$-model action, it is convenient to treat the system in a more general background and follow the approach used in \cite{Kruczenski:2004cn} generalized to the case where also an antisymmetric field is included.
The procedure is the following.
We start from a Polyakov action describing the $\sigma$-model action corresponding to a generic background \eqref{general_metric} including
 also an antisymmetric field $B$. The Lagrangian is first T-dualized along a longitudinal coordinate $\xi$ of the deformed sphere $\sphere^5$, the Nambu-Goto action is then obtained by eliminating the world sheet metric, and finally, a large $J$ limit is taken.

The general form of the ten-dimensional string background metric is the following
\be
\label{general_metric}
ds^2=R^2\left(-f(x^i)ds^2_{AdS}+G_{ij}(x^i) dx^idx^j+(H(x^i))^{-1} d\xi d \xi +2 d \xi C \right)\,.
\ee
The target space coordinates $X^M$ with $M,N=0,\dots, 9$ are $X^M=(t, y^{i'},x^i,\xi)$, with $(t,y^{i'})$ the coordinates 
in $\ads_5$, $i'=1,\dots,4$, and with the coordinates $(x^i,\xi)$ in $\sphere^5$, $i=1,\dots,4$. $C$ is a one form, i.e. 
$C=C_i dx^i$, and $R$ is the AdS curvature radius. We also demand that the B-field
should be of the form
\be
\label{Btilde}
B=R^2(\tilde{B}_{i} dx^{i}\wedge d \xi +B_{ij} dx^i\wedge dx^j)\,.
\ee
The bosonic string action is 
\be
\label{def_bosnicaction_v1}
S&=&\sqrt{\lambda} \int d\tau \int_0^{2\pi} \frac{d\sigma}{2\pi} \mathcal{L} \,
\nn\\
& =& - {1\over 2 \alpha'} \int d\tau \int_0^{2\pi} {d\sigma \over 2\pi} 
\left [ \sqrt{-h}h^{\alpha\beta}\p_\alpha X^M \p_\beta X^N G_{MN}
        -\epsilon^{\alpha\beta} \p_\alpha X^M \p_\beta X^N B_{MN}
        \right] 
        \qquad
\ee
where the world-sheet coordinates $\sigma^\alpha= (\tau\,, \sigma)$ are labelled by greek indices $\alpha,\, \beta=0,1$, the Latin capital 
letters $M,N=0,\dots, 9$ label the curved target space indices, and $h^{\alpha\beta}$, $\epsilon^{\alpha\beta}$ are the world-sheet metric 
and the totally 2d antisymmetric tensor respectively. 
In \eqref{def_bosnicaction_v1} we have also used the $\mbox{AdS}_5/ \mbox{CFT}_4$ relation among the 't Hooft coupling $\lambda$ and the radius
 curvature $R$, i.e.
$
\sqrt{\lambda}=\frac{R^2}{\alpha'}
$.
For the specific form of the metric (\ref{general_metric}) the Lagrangian is
\be
\label{eq1}
\mathcal{L} &= &-\frac1{2}\sqrt{-h}h^{\alpha \beta}(-f \partial_{\alpha} t\partial_{\beta} t
+G_{ij}\partial_{\alpha} x^i \partial_{\beta} x^j +
H^{-1}\partial_{\alpha} \xi \partial_{\beta} \xi+2\partial_{\alpha} \xi C_{\beta})\nn \\
&+&\epsilon^{\alpha \beta} ( \tilde{B}_{\alpha} \partial_{\beta} \xi
+\frac{1}{2} B_{ij}\partial_{\alpha} x^i \partial_{\beta} x^j)\,,
\ee
with $
\tilde{B}_{\alpha}=\tilde{B}_i \partial_{\alpha} x^i$, $C_{\alpha}=C_i\partial_{\alpha} x^i$. We omit the dependence of the background fields 
on the coordinates. 
The equation of motion for the coordinate $\xi$ is
\be
\label{eom_alpha}
\partial_{\alpha} \left( \sqrt{-h} h^{\alpha \beta} (H^{-1}\partial_{\beta} \xi +C_{\beta}) +\epsilon^{\alpha \beta} \tilde{B}_{\beta} \right)=0\,.
\ee
This relation can be automatically satisfied by defining $\tilde{\xi}$ as
\be
\label{above_eq}
  \sqrt{-h} h^{\alpha \beta} (H^{-1}\partial_{\beta} \xi +C_{\beta}) +\epsilon^{\alpha \beta} \tilde{B}_{\beta} =
-\epsilon^{\alpha \beta}\partial_{\beta}\tilde{\xi}\,.
\ee
The negative sign in the above equation \eqref{above_eq} is chosen for later convenience. Finally, the conserved 
charge coming from translation invariance along $\xi$ direction is \cite{Kruczenski:2004cn}
\be\label{J}
\mathcal{J}=\int_0^{2\pi}\frac{d\sigma}{2\pi} P_{0} \,,
\qquad P^{\alpha}=\frac{\delta \mathcal{L}}{\delta (\p_\alpha \xi)}=\sqrt{-h} h^{\alpha \beta} (H^{-1}\partial_{\beta} \xi +C_{\beta}) 
+\epsilon^{\alpha \beta} \tilde{B}_{\beta}\,.
\ee
From the definition of $\tilde \xi$ \eqref{above_eq} and the expression for the conserved charge \eqref{J}, it follows the relation $\tilde{\xi}=\mathcal{J}\sigma$, which we use later.

\paragraph{T-dualizing.} 

The longitudinal coordinate $\xi$ is redundant (as well as $t$) and we can integrate it out by a T-duality%
\footnote{For a review on T-dualization cf. {\it e.g.} \cite{Alvarez:1994dn}.}
 ({\it i.e.} a 2d-duality). This amounts to substitute the generalized momentum $\p_\alpha \xi$ 
 with an auxiliary field $A_{\alpha}$ and by contemporarily introducing a Lagrange multiplier
term, $-\epsilon^{\alpha \beta} A_{\alpha}\partial_{\beta}\tilde{\xi}$, such that the equation of motion \eqref{eom_alpha} holds. 
Thus, the Lagrangian \eqref{eq1} becomes
\be
\label{eq2}
\mathcal{L} &=&-\frac1{2}\sqrt{-h}h^{\alpha \beta}(-f\partial_{\alpha} t\partial_{\beta} t
+G^{ij}\partial_{\alpha} x_i \partial_{\beta} x_j +
H^{-1}A_{\alpha} A_{\beta} +2 A_{\alpha}  C_{\beta})\nn \\
&-&\epsilon^{\alpha \beta} (A_{\alpha} \tilde{B}_{\beta}
+ A_{\alpha}\partial_{\beta}\tilde{\xi}
-\frac{1}{2} B_{ij}\partial_{\alpha} x^i\partial_{\beta} x^j)\,.
\ee
The equation of motion coming from the constraint is
\be
\sqrt{-h}h^{\alpha \beta}(H^{-1}A_{\beta}+C_{\beta})+\epsilon^{\alpha \beta} \tilde{B}_{\beta}=-\epsilon^{\alpha \beta}\partial_{\beta}\tilde{\xi}\,,
\ee
which implies that $A$ is defined as
\be
\label{A:constraint}
A_{\gamma}=-H(x^i) \left(C_{\gamma}+ {h_{\gamma \beta}\over \sqrt{-h}}\epsilon^{\beta \alpha}(\tilde{B}_{\alpha}+\partial_{\alpha}\tilde{\xi}) \right)\,.
\ee
The Lagrangian \eqref{eq2} in terms of the expression (\ref{A:constraint}) for $A_\alpha$ is
\be
\mathcal{L} &=&-\frac1{2}\sqrt{-h}h^{\alpha \beta}(-f \partial_{\alpha} t\partial_{\beta} t
+G_{ij}\partial_{\alpha} x^i \partial_{\beta} x^j - 
HC_{\alpha} C_{\beta} 
+H(\tilde{B}_{\alpha}+\partial_{\alpha} \tilde{\xi})(\tilde{B}_{\beta}+\partial_{\beta}\tilde{\xi}))
\nn \\ 
&-&\epsilon^{\alpha \beta} (C_{\alpha}(\tilde{B}_{\beta}+\partial_{\beta}\tilde{\xi})-
\frac{1}{2} B_{ij}\partial_{\alpha} x^i\partial_{\beta} x^j)\,.
\ee
%

\paragraph{The Nambu-Goto action.} 

In order to obtain the Nambu-Goto action we need to eliminate the two dimensional world-sheet metric $h^{\alpha \beta}$ according to
\be
\label{Nambu-Goto}
\mathcal{L}=-\epsilon^ {\alpha \beta} \left(C_{\alpha} (B_{\beta}+\partial_{\beta} \tilde{\xi}) H-
\frac{1}{2} B_{ij}\partial_{\alpha} x^i\partial_{\beta} x^j\right)
-\sqrt{\det_{\alpha\beta} g_{\alpha\beta}}\,,
\ee
where $g_{\alpha \beta}$ is the embedding world-sheet metric given by
\be
\label{Embedded_metric}
g_{\alpha \beta}=-f \partial_{\alpha} t\partial_{\beta} t+ G_{ij} \partial_{\alpha} x^i\partial_{\beta} x^j
-C_{\alpha}C_{\beta}H+(B_{\alpha}+\partial_{\alpha}\tilde{\xi})(B_{\beta}+\partial_{\beta}\tilde{\xi}) H\,.
\ee

\paragraph{The static gauge.} 

Finally, we gauge-fix the longitudinal coordinate $t$ by choosing the static gauge
\be\label{static_gauge_true}
t=\tau\,,
\ee
with $\tau$ the world-sheet time coordinate. Using \eqref{static_gauge_true} 
and the relation $\tilde{\xi}=\mathcal{J} \sigma$,%
\footnote{The relation $\tilde{\xi}=\mathcal{J} \sigma$ ensures that the generalized momentum is uniformly distributed along $\sigma$.}
the first contribution to the Lagrangian \eqref{Nambu-Goto}, namely the Wess-Zumino (WZ) term, becomes
\be
\label{L_wz}
\mathcal{L}_{1}
\equiv -\epsilon^ {\alpha \beta} C_{\alpha} (B_{\beta}+\partial_{\beta} \tilde{\xi}) H
=-(C_0 \mathcal{J} +\epsilon^{\alpha \beta} C_{\alpha} \tilde{B}_{\beta})H\,.
\ee
The second term of \eqref{Nambu-Goto} can be ignored because it does not contribute after the fast spinning string limit
has been taken. Finally, the determinant of the embedding metric in the static gauge \eqref{static_gauge_true} is given by
\be
\label{det_g}
\det_{\alpha,\beta}{g_{\alpha \beta}} &= & (-f +G_{ij}\partial_0 x^i \partial_0 x^j -(C_0^2-\tilde{B}_0^2)H )
(G_{ij}\partial_1 x^i \partial_1 x^j -(C_1^2-(\tilde{B}_1+\mathcal{J})^2)H)
\nn\\ &-&(G_{ij}\partial_1 x^i \partial_0 x^j-(C_1C_0-\tilde{B}_0(\tilde{B}_1+\mathcal{J}))H)^2\,.
\ee

\paragraph{The fast string expansion.} 

In order to have a consistent expansion in $1/\mathcal{J}$, we rescale the world-sheet time coordinate, $\tau\rightarrow \mathcal{J}^2\tau$, which implies 
$\partial_0 \rightarrow 1/\mathcal{J}^2 \; \partial_0$. The parameters are related according to
\be
J=\sqrt{\lambda}\mathcal{J}\,, \qquad \lmtilde =\frac{1}{\mathcal{J}^2}\,.
\ee
The WZ term of the Lagrangian \eqref{L_wz} becomes
\be\label{Lwzw}
\mathcal{L}_{1}=\left(\frac{1}{ \mathcal{J}} C_0 +\frac1{\mathcal{J}^2}\epsilon^{\alpha \beta} C_{\alpha} \tilde{B}_{\beta}\right) H \,.
\ee
By expanding the determinant of the embedding metric $g_{\alpha\beta}$ \eqref{det_g} for large values of $\mathcal J$ and by assuming that in \eqref{det_g} no 
functions grow faster than $\mathcal{J}^0$, we obtain
\be
\det_{\alpha,\beta} {g_{\alpha \beta}} =-f \mathcal{J}^2\left(H \left(1+\frac{2}{\mathcal{J}}\tilde{B}_1\right) +\frac1{\mathcal{J}^2}\left(H \tilde{B}_1^2-
C_1^2+G_{ij}\partial_1 x^i \partial_1 x^j\right)\right) \,,
\ee
and the terms can be expanded up to second order in a small parameter $\smtilde$ ($\smtilde \sim \mathcal{J}^{-1}$)
 as
\be
&&f = (1+\smtilde^2 f^{(2)}) \,, \qquad
C=C^{(0)}+\smtilde^2 C^{(2)} \,, \qquad
\tilde{B}=\smtilde \tilde{B}^{(1)}\,, \nn \\
&& H=1+\smtilde^2 H^{(2)} \, , \qquad
G_{ij}=G_{ij}^{(0)}+\smtilde^2 G_{ij}^{(2)}\,.
\ee
%
%
Hence, the square root of the determinant is given by
\be
\label{det1}
\sqrt{\det_{\alpha,\beta}{g_{\alpha \beta}}} &=& \mathcal{J} \sqrt{|f |}\left(1+
\frac1{\mathcal{J}^2}(\mathcal{J}\smtilde \tilde{B}_1^{(1)}+\frac1{2}(\mathcal{J}^2\smtilde^2 H^{(2)}
 -(C_1^{(0)})^2+G_{ij}^{(0)}\partial_1 x^i \partial_1 x^j))\right)\nn \\
&=&  \mathcal{J} 
+
\frac1{\mathcal{J}}\left(\mathcal{J}\smtilde \tilde{B}_1^{(1)}+\frac1{2}(\mathcal{J}^2\smtilde^2 H^{(2)}
 -(C_1^{(0)})^2+G_{ij}^{(0)}\partial_1 x^i \partial_1 x^j) 
 + \frac{\smtilde^2 \mathcal{J}^2}{2} f^{(2)}\right) \,.
 \nn \\
\ee
Finally, by combining the WZ term \eqref{Lwzw} and the above expression \eqref{det1}, the Lagrangian takes the form
\be
\label{Lagrangian: General}
\mathcal{L} &=& \mathcal{J} +
\frac1{\mathcal{J}} C^{(0)}_0\nn\\
& -&
\frac1{\mathcal{J}} \left(\mathcal{J}\smtilde \tilde{B}_1^{(1)}+\frac1{2}(\mathcal{J}^2\smtilde^2 H^{(2)}
 -(C_1^{(0)})^2+G_{ij}^{(0)}\partial_1 x^i \partial_1 x^j)
 + \frac{\smtilde^2 \mathcal{J}^2}{2} f^{(2)} 
\right)\,.~
\ee
By defining $\mathcal{\tilde{L}}$ as 
$
\mathcal{L}\, \equiv \, \frac1{\mathcal{J}}\, \mathcal{\tilde{L}}$ and $\bar{\sigma}\equiv \,\mathcal{J}\, \smtilde\,
$,
the action is written as
\be
S=J\int d\tau \int_0^{2\pi}d\sigma \mathcal{\tilde{L}}\,,
\ee 
with
\be
\label{Lagrangian:General2}
\mathcal{\tilde{L}}=\mathcal{J}^2+
 C^{(0)}_0 -(
\bar{\sigma} \tilde{B}_1^{(1)}+\frac1{2}(\bar{\sigma}^2 H^{(2)}
 -(C_1^{(0)})^2+G_{ij}^{(0)}\partial_1 x^i \partial_1 x^j) 
 + \frac{\bar{\sigma}^2}{2} f^{(2)})\,.
\ee
This is the formula which we will use in the next section in order to obtain the fast string action for
the mixed sector in the Lunin-Maldacena background.

\subsection{The Lunin-Maldacena background}
\label{sec:applyLM}

In order to fix the notation we recall the Lunin-Maldacena (LM) background for complex $\beta$ values. Such a background is a
non-trivial modification of the metric derived for the real $\beta$ case, since also S-dualities are involved in its construction~\cite{Lunin:2005jy}. 
More details are given in appendix \ref{sec:geom_setup}.
We use the convention%
\footnote{Note that our convention differs from the one adopted in~\cite{Lunin:2005jy}, where they instead have $q=e^{-i\beta}$ in
the superpotential.} 
$q=e^{i2\pi \beta}$ 
and then $\beta=\gamma-i\sigma$. In this paper we consider only the case where 
$
\tilde\gamma\equiv R^2 \gamma=0
$.

For this specific case the Lunin-Maldacena metric \cite{Lunin:2005jy} (cf. also~\cite{Frolov:2005ty}) reduces to
\be
\label{metric:gamma=0}
&& ds^2= R^2 \,G^{-1/2} \left[ 
            ds^2_{\ads_5}
            +\sum_{i=1}^3 (d\rho_i^2+G\rho_i^2 d\phi_i^2)
            +\tilde\sigma^2 G \rho_1^2\rho_2^2\rho_3^2 (\sum_{i=1}^3 d\phi_i)^2\right]\,,
 \nn \\
&& 
 G^{-1} \equiv 1+\tilde\sigma^2 Q\,,
 \qquad
 Q \equiv \rho_1^2\rho_2^2+\rho_2^2\rho_3^2+\rho_1^2\rho_3^2\,,
\ee
where the deformation parameter is
\be
\tilde\sigma\equiv \sigma R^2 \,.
\ee
The B-field for $\tilde\gamma=0$ is 
\be\label{eq:oneforms0}
&& B  =  12 R^2 \,\tilde\sigma\, w_1 \,d\psi \,, \quad
\nn \\
&& d w_1 \equiv \cos \theta \sin^3 \theta \sin\psi \cos\psi d\theta \wedge d\psi \,, \quad
 \psi = {1\over 3} \left ( \phi_1 +\phi_2+\phi_3\right)\,.
\ee 

In order to consider the mixed $\grSU_q(3)$ sector, a slightly different parameterization has to be chosen with respect to the standard one commonly used in literature, {\it e.g.} cf. \cite{Frolov:2005ty}, namely 
\be
\label{parameterization}
&& \phi_1 = \xi +\varphi_1 \,, \qquad \phi_2 = -\xi +\varphi_1 \,, \qquad \phi_3= \xi +\varphi_2 \,,
\nn \\
&& \rho_1= \sin\th \cos\psi \,, \qquad \rho_2 = \sin\th \sin\psi \,, \qquad \rho_3= \cos\th \,.
\ee
The main difference between the mixed $\grSU_q(3)$ sector and the purely holomorphic  $\grSU_q(3)$ sector is the orientation of the coordinate $\phi_2$ with respect to the slow parameter
$\xi$. 
We stress once more that for the non-deformed theory this would not produce any difference. In fact, in section \eqref{sec:compare_holomorphic} we will see that the only difference between the mixed and purely holomorphic $\grSU_q(3)$ sector comes from terms multiplied by the deformation parameter $\tilde\sigma$, which is obviously absent in the non-deformed SYM.

Let us now apply the formula \eqref{Lagrangian:General2} derived in the previous section on the background \eqref{metric:gamma=0} using the parameterization \eqref{parameterization}. The metric components are
\be
\label{functions:mixed}
&& G^{(0)}_{\varphi_1 \varphi_1}=\sin^2 \theta\,, \qquad G^{(0)}_{\varphi_1 \varphi_1}= \cos^2 \theta
\,,\qquad
 G^{(0)}_{\theta \theta}=1\,, \qquad G^{(0)}_{\psi \psi}=\sin^2 \theta \,,
\nn \\
&& C^{(0)}= \left(\sin^2 \theta \cos 2\psi d{\varphi}_1+\cos^2 \theta d{\varphi}_2 \right)\,,   \nn \\
&& H(x_i)^{-1}= G^{1/2} \left(1+\rho_1^2\rho_2^2\rho_3^2 \right)= \left(1+\smtilde^2 \left(\rho_1^2\rho_2^2\rho_3^2-\frac{Q}{2} \right)\right)+\mathcal{O}(\smtilde^3)\nn \,, \\
&& f(x_i)=\left(1+\smtilde^2 Q\right)^{1/2}=\left(1+\frac{\smtilde^2}{2} Q\right)+\mathcal{O}(\smtilde^3) \,.
\ee
Notice that there is an ambiguity in writing the antisymmetric B-field, which can be thought as a gauge choice, in fact $d w_1$ can be chosen in different ways according to our parametrization \eqref{parameterization}, \emph{i.e.}
\be
\label{dw1}
dw_1=\frac1{4} d(\rho_1^2)\wedge d(\rho_2^2)=\frac1{4} d(\rho_2^2)\wedge d(\rho_3^2)=\frac1{4} d(\rho_3^2)\wedge d(\rho_1^2)\,.
\ee
In appendix \ref{sec:coherent_sigma_mod}, in the coherent state spin chain approach, we will see that the terms corresponding to the different gauge choices of the antisymmetric B-field indeed cancel for a closed spin chain, and they do not affect the physics of the model, as we expected indeed.
We choose $w_1$ to be of the form
%
\be
\label{B:mixed gauge}
4w_1=\rho_{3}^2\rho_{1} d\rho_{1}-\rho_{1}^2\rho_{3} d\rho_{3}\,,
\ee
which is directly related to the
expression of the spin-chain Hamiltonian for the $\grSU_q(3)$ sector. 
This choice gives the same linear term which appears from
the coherent sigma model derived from the spin chain Hamiltonian \eqref{spinchain_h}.
For the mixed sector we are discussing and with the parametrization \eqref{parameterization}, we have
\be
\label{eqb}
B=4\smtilde w_1 ( d\xi+2 d\varphi_1+d\varphi_2 )\,.
\ee
Notice that only the term $\tilde{B}$ defined in equation (\ref{Btilde}) contributes
\be
\label{bfield}
\tilde{B}&=&4\smtilde w_1 
=(\rho_{3}^2\rho_{1} d\rho_{1}-\rho_{1}^2\rho_{3} d\rho_{3})
\nn \\
&=&\cos \theta\sin \theta \cos^2 \psi \,d\theta
-\cos^2\theta \sin^2\theta \cos \psi \sin \psi \, d\psi\,,
\ee
with the gauge choice (\ref{B:mixed gauge})\,.
Inserting the expressions \eqref{functions:mixed} and \eqref{bfield} into the Lagrangian (\ref{Lagrangian:General2}), we obtain
\be
\label{Lagrangian:mixed}
\mathcal{L} &=& \sin^2 \theta \cos 2\psi \dot{\varphi}_1+\cos^2 \theta \dot{\varphi}_2-
\frac{\bar{\sigma}^2 }{2} \sin^2 \theta(\cos^2 \theta +\frac1{4}\sin^4\theta 
\sin^2 2\psi)
\nn \\ &+&\frac1{2}(\sin^2\theta \cos 2\psi \,\varphi_1'+\cos^2\theta \,\varphi_2')^2
-\frac1{2}({\theta'}^2 +\sin^2 \theta \,{\psi'}^2+\sin^2 \theta \,{\varphi_1'}^2+\cos^2\theta \,{\varphi_2'}^2)\nn \\
& - &\bar{\sigma}  (\sin \theta \cos \theta \cos^2 \psi \,\theta'-\sin^2\theta\cos^2\theta\cos \psi \sin \psi \,\psi')\,.
\ee
%
Up to an additional term $\dot{\xi}$, whose only effect is to make the conservation of the charge $J$ manifest already at this level~\cite{Kristjansen:2004za}, this leads to the following action:
\be
\label{action_string2}
S=J\int d\tau \int_0^{2\pi}d\sigma \left(\rho_1^2\dot{\phi}_1-\rho_2^2\dot{\phi}_2+\rho_3^2 \dot{\phi}_3-\frac1{2} \mathcal{H}\right)\,,
\ee
with the Hamiltonian given by
\be
\label{Hamiltonian:mixed}
\mathcal{H}&=&
(\rho_1 \rho'_2-\rho_2 \rho'_1)^2
+(\rho_2 \rho'_3-\rho_3 \rho'_2)^2
+(\rho_3 \rho'_1-\rho_1 \rho'_3) (\rho_3 \rho'_1-\rho_1 \rho'_3+2\bar{\sigma} \rho_1\rho_3)
 \nn\\
&+&\bar{\sigma}^2 (\rho_1^2 \rho_2^2 +\rho_1^2 \rho_3^2 +\rho_3^2 \rho_2^2-\rho_1^2 \rho_2^2 \rho_3^2)
\nn\\
&+& (\phi'_1+\phi'_2)^2\rho_1^2 \rho_2^2 
+  (\phi'_1-\phi'_3)^2\rho_1^2 \rho_3^2
+ (\phi'_2+\phi'_3)^2\rho_2^2 \rho_3^2\,.
\ee
%
By undoing the rescaling of the time components, namely $\tau\rightarrow \lmtilde \tau\equiv 1/\mathcal{J}^2 \tau$, the Lagrangian becomes
\be
\label{LA}
\mathcal{L}=\rho_1^2\dot{\phi}_1-\rho_2^2\dot{\phi}_2+\rho_3^2 \dot{\phi}_3-\frac1{2}\lmtilde \mathcal{H}\,,
\ee
or explicitly in terms of the parameterization \eqref{parameterization} 
\be
\label{L_string_final}
\mathcal{L} &=& \dot{\xi}+ \sin^2 \theta \cos 2\psi \dot{\varphi}_1+\cos^2 \theta \dot{\varphi}_2 -\frac{\lmtilde }{2}\Big\{
\bar{\sigma}^2  (\sin^2 \theta(\cos^2 \theta +\frac1{4}\sin^4\theta 
\sin^2 2\psi))
\nn \\ &-&(\sin^2\theta \cos 2\psi \varphi_1'+\cos^2\theta \varphi_2')^2
+({\theta'}^2 +\sin^2 \theta {\psi'}^2+\sin^2 \theta {\varphi_1'}^2+\cos^2\theta\varphi_2)\nn \\
 & +& 2\bar{\sigma} (\sin \theta \cos \theta \cos^2 \psi\, \theta'-\sin^2\theta\,\cos^2\theta\cos \psi \sin \psi\, \psi')\Big\}\,.
\ee 


\subsection{Circular string solution}
\label{subsec:circular_string_sol_v2}

Following \cite{Frolov:2005ty, Kristjansen:2004za, Hernandez:2004uw} one can search for classical circular string solutions such as
\be
\label{circular_sol_v1}
\theta=\theta_0 \,, \qquad
\psi=\psi_0\,, \qquad
\varphi_1' =m\,, 
\qquad
\varphi_2'=n\,,
\ee
where $m$ and $n$ are integers.
%
%
%
%
%
%
The conserved charges
\be
\label{charges}
&& P_{\varphi_1}=\int_0^{2\pi} {d\sigma\over 2\pi} {\delta \mathcal L\over \delta \dot\varphi_1}
=  J \int_0^{2\pi} {d\sigma\over 2\pi} \cos{2\psi_0}\sin^2\theta_0=J(\rho_1^2-\rho_2^2)\,,
\nn \\
&& P_{\varphi_2}=\int_0^{2\pi} {d\sigma\over 2\pi} {\delta \mathcal L\over \delta \dot\varphi_2}
= J  \int_0^{2\pi}  {d\sigma\over 2\pi} \cos^2\theta_0=J\rho_3^2\,,
\ee
can be identified with $J_3=P_{\varphi_2}$ and $P_{\varphi_1}=J_1-J_2$. By using the following relations%
%
%
\be
\rho_1^2=j_1\,, \qquad \rho_2^2=j_2 \,,\qquad \rho_3^2=j_3 \qquad \mbox{where} \qquad j_i\equiv \frac{J_i}{J}\,,
\ee
together with the equations of motion, the Hamiltonian \eqref{Hamiltonian:mixed} reads
\be
\label{Energy:string}
\mathcal{H}&=
&\frac{\lambda}{2 J} \Big(4m^2j_1 j_2 
+  (m-n)^2j_1 j_3
+ (m+n)^2 j_2 j_3
\nn \\
&+&\bar{\sigma}^2 (j_1 j_2 +j_1 j_3 +j_3 j_2-j_1 j_2 j_3) \Big)\,.
\ee
This can be compared with the resulting Hamiltonian for the holomorphic deformed $\grSU_q(3)$ sector, where there would have been 
a $9$ in front of the last cubic term~\cite{Frolov:2005iq}. Notice that for rational
string solutions, the linear terms in the Lagrangian \eqref{L_string_final} do not contribute to the energy.




\subsection{A comparison with the holomorphic $\grSU_q(3)$ sector}
\label{sec:compare_holomorphic}

The Lagrangian for the three holomorphic state sector was first derived in \cite{Frolov:2005ty}. Here we re-derive it by using
the expression \eqref{Lagrangian:General2} and our convention for reader's convenience.
For the purely holomorphic sector the Lagrangian is
\be
\label{LA_h}
\mathcal{L}=\rho_1^2\dot{\phi}_1+\rho_2^2\dot{\phi}_2+\rho_3^2 \dot{\phi}_3-\frac1{2}\lmtilde \mathcal{H}\,,
\ee
with 
\be
\label{Hamiltonian:holomorphic}
\mathcal{H}&=&
(\rho_1 \rho'_2-\rho_2 \rho'_1+\bar\sigma \rho_1\rho_2)^2
+(\rho_2 \rho'_3-\rho_3 \rho'_2+\bar\sigma \rho_2\rho_3)^2
+(\rho_3 \rho'_1-\rho_1 \rho'_3+\bar\sigma \rho_1\rho_3)^2
 \nn \\
&-&9\bar\sigma^2 \rho_1^2 \rho_2^2 \rho_3^2
+ (\phi'_1-\phi'_2)^2\rho_1^2 \rho_2^2 
+  (\phi'_1-\phi'_3)^2\rho_1^2 \rho_3^2
+ (\phi'_2-\phi'_3)^2\rho_2^2 \rho_3^2\,,
\ee
which in terms of the following parameterization
\be
\label{holomorphic_angles}
&& \phi_1 = \xi+\varphi_1 \,, \qquad \phi_2 = \xi -\varphi_1 \,, \qquad \phi_3= \xi +\varphi_2 \,,
 \nn \\
&& \rho_1= \sin\th \cos\psi \,, \qquad \rho_2 = \sin\th \sin\psi \,, \qquad \rho_3= \cos\th \,,
\ee
becomes
\be
\mathcal{L} &= &(\sin^2 \theta \cos 2\psi \dot{\varphi}_1+\cos^2 \theta \dot{\varphi}_2)
\nn \\
&-&\frac{\lmtilde }{2}(
\bar{\sigma}^2  \sin^2\theta(\cos^2 \theta -{1\over 4} (8\cos^2\theta-\sin^2\theta)\sin^2\theta \sin^2{2\psi})
\nn\\ &-&(\cos^2\theta \varphi_2' -\sin^2\theta \cos 2\psi \varphi_1')^2
+({\theta'}^2 +\sin^2 \theta {\psi'}^2+\sin^2 \theta {\varphi_1'}^2+\cos^2\theta\varphi_2)\nn\\
&+& 2\bar\sigma  (\sin \theta \cos \theta \cos^2 \psi \theta'+\sin^2\theta (1-3\cos^2\theta) \cos \psi \sin \psi \psi'))\,.
\ee 
Notice the change of sign between the angles parameterizing the deformed compact space in the two cases, cf. \eqref{parameterization} and \eqref{holomorphic_angles}, as well in the Lagrangian \eqref{LA} and \eqref{LA_h}. 
%

We complete the comparison between the purely holomorphic and the mixed deformed $\grSU_q(3)$ sector%
\footnote{For both holomorphic and mixed sectors a quantum deformation of the algebra of functions on $\grSU(3)$
appears (which we denote $\grSU_q(3)$), but they are not the same. For the mixed sector it is the standard quasi-triangular 
Hopf algebra, while for the holomorphic sector it is a non quasi-triangular Hopf algebra  \cite{Mansson:2008xv}.}
by re-expressing the background fields \eqref{metric:gamma=0}  and \eqref{eq:oneforms0} in terms of the new parameterization \eqref{holomorphic_angles}. 
It is straightforward to see that there are only two differences among the two types of $\grSU_q(3)$ sectors. The first difference comes from the background metric and it is quadratic in the small deformation parameter $\tilde\sigma$, while the other one is due to the B-field and linear in $\tilde\sigma$.
We read off the functions $f(x_i)$, $H(x_i)$ and $G_{ij}(x^i)$ from the metric
\be
\label{holomorphic:metriccomp}
&&G^{(0)}_{\varphi_1 \varphi_1}= \sin^2 \theta\,, \qquad G^{(0)}_{\varphi_1 \varphi_1}= \cos^2 \theta\,,
\qquad
C^{(0)}=(\sin^2 \theta \cos 2\psi d{\varphi}_1+\cos^2 \theta d{\varphi}_2)\,,\nn  \\
&& H = G^{1/2}(1+9 \rho_1^2\rho_2^2\rho_3^2)=R^2(1+\smtilde^2 (9 \rho_1^2\rho_2^2\rho_3^2-\frac{Q}{2}))+\mathcal{O}(\smtilde^3) \,,\nn \\
&& f =(1+\smtilde^2 Q)^{1/2}=(1+\frac{\smtilde^2}{2} Q)+\mathcal{O}(\smtilde^3)\,.
\ee
These expressions can be compared with equation (\ref{functions:mixed}). Up to one-loop order in $\lmtilde$, the only difference is in the quadratic contribution in $\smtilde$ of the $H(x_i)$ term, in particular
only the coefficient in front of $\rho_1^2\rho_2^2\rho_3^2$ differs.
Concerning the linear term, the only contributing difference comes from the antisymmetric field. A convenient gauge choice for the antisymmetric field $B$ turns out to be
\be
\label{w_hol}
4w_1=\frac1{3}\sum_{i=1}^3 (\rho_{i}^2\rho_{i+1} d\rho_{i+1}-\rho_{i+1}^2\rho_{i} d\rho_{i})\,,
\ee
where we assume a cyclicity in the indices. With this gauge choice and with the parameterization \eqref{holomorphic_angles}, the antisymmetric field is
\be
\label{eqb2}
B=4\smtilde w_1 (3 d\xi+d\varphi_2 )\,, \qquad \tilde{B}=12\smtilde w_1\,,
\ee
where $\tilde{B}$ is defined in (\ref{Btilde}).
The first term in equation \eqref{eqb2} can be compared with the result for the mixed sector \eqref{eqb}:
They only differ by a factor of three, which reflects the fact that we have only one of the three linear deformed terms present in the mixed $\grSU_q(3)$ sector.

\section{Solving Bethe equations}
\label{sec:BE}
In this section we show that the Bethe equations coming from the one-loop dilatation operator for the mixed $\grSU_q(3)$ sector~\cite{Mansson:2007sh} give the same
energy obtained from the fast spinning string action \eqref{Lagrangian:mixed} for the rational string solution~\eqref{circular_sol_v1}. The conserved currents $J_i$ correspond to the number of fields $\phi_i$ on the gauge theory side. We are considering operators of the form $\mathrm{Tr}( \phi_1^{J_1}\,\bar{\phi}_2^{J_2}\,\phi_3^{J_3})$.
The length of the spin chain is then $J=J_1+J_2+J_3$.
As for the undeformed $\grSU(3)$ spin chain, there are two sets of Bethe roots $\{u_{1,j}\}_{j=1}^{K_1}$ and $\{u_{2,j}\}_{j=1}^{K_2}$. The relations
between the roots and the currents $J_i$ are the same as in the undeformed $\grSU(3)$ spin chain case, (cf. for instance \cite{Freyhult:2005fn}), namely
\be
(J_1,J_2,J_3)=(J-K_1, K_1-K_2, K_2)\,.
\ee
The set of
algebraic Bethe equations for the mixed $\grSU_q(3)$ sector have been derived in \cite{Mansson:2007sh}. Here, we recollect them for completeness 
\begin{equation}
\label{bethe1}
 \prod_{l\neq k}^{K_2} \frac{\sinh((\mu_{2,k}-\mu_{2,l})-2\pi\sigma)}{\sinh((\mu_{2,k}-\mu_{2,l})+2\pi\sigma)}\prod_{j=1}^{K_1} 
\frac{\sinh((\mu_{2,k}-\mu_{1,j})+\pi\sigma)}{\sinh ((\mu_{2,k}-\mu_{1,j})-\pi\sigma)}=1\,,
\end{equation}
\be
\label{bethe2}
 \left(
\frac{\sinh (\mu_{1,k}+\pi\sigma)}{\sinh(\mu_{1,k}-\pi\sigma)}\right)^J=
\prod_{l\neq k}^{K_1} \frac{\sinh((\mu_{1,k}-\mu_{1,l})+2\pi\sigma)}{\sinh((\mu_{1,k}-\mu_{1,l})-2\pi\sigma)}\,
\prod_{j=1}^{K_2} \frac{\sinh((\mu_{1,k}-\mu_{2,j})-\pi\sigma)}{\sinh((\mu_{1,k}-\mu_{2,j})+\pi\sigma)}\,,
\quad \ee
and we have also set the
phase contribution of $q$ to zero, namely $\gamma=0$.
The cyclicity constraint is
\begin{equation}
 \prod_{l}^{K_1} \frac{\sinh(\mu_{1,l}+\pi\sigma)}{\sinh(\mu_{1,l}-\pi\sigma)}=1\,,
\end{equation}
and the energy is given by
\begin{equation}
\label{bethe_energy1}
E=\sum_{k=1}^{K_1}\epsilon_k, \qquad \tn{with} \qquad 
\epsilon_k=\frac{\lambda}{8\pi^2}
\frac{\sinh^2 2\pi\sigma}{\sinh(\mu_{1,k}-\pi\sigma)\sinh(\mu_{1,k}+\pi\sigma)}\,.
\end{equation} 
Since we want to compare the Bethe equations with a fast string configuration, we have to consider very long spin chain, {\it i.e.} very long operators, which implies taking the limit where $J\rightarrow \infty$, $\sigma\rightarrow 0$ and the product $\sigma J=\bar{\sigma}$ is finite. 
Taking the logarithm of the equations \eqref{bethe1} and \eqref{bethe2} and defining $x_{2,k}$ through $\tanh \mu_{2,k}=2i\tanh (\pi\sigma) J x_{2,k}$, in the small $\sigma$ expansion the expressions \eqref{bethe1}-\eqref{bethe2} become%
\footnote{Further details are given in Appendix \ref{subsec:Solving the Bethe equations}.}
\begin{equation}
\label{eq:1}
2\pi m_k=\frac{2}{J} \sum_{l\neq k}^{K_2}
\frac{1+(2\pi\bar{\sigma} )^2x_{2,k}x_{2,l}}{x_{2,k}-x_{2,l}}
-\frac{1}{J}\sum_{j=1}^{K_1} 
\frac{1+(2\pi \bar{\sigma} )^2x_{2,k}x_{1,j}}{x_{2,k}-x_{1,j}} \,,
\end{equation}
\bea
\label{eq:2}
 &&\frac{1}{x_{1,k}}+2\pi n_k=
-\frac{2}{J}\sum_{l\neq k}^{K_1} 
\frac{1+({2\pi\bar{\sigma}} )^2x_{1,k}x_{1,l}}{x_{1,k}-x_{1,l}}
\,
+\frac{1}{J}\sum_{j=1}^{K_2} 
\frac{1+(2\pi\bar{\sigma} )^2x_{2,k}x_{1,j}}{x_{1,j}-x_{2,k}}\,.
\eea
In this limit we obtain for the energy
\be
E= \frac{\lambda}{8\pi^2} \left(\sum_k^{K_1}\frac1{x_{1,k}^2}+(2\pi\bar{\sigma} )^2K_1 \right)\,,
\ee
and the cyclicity constraint becomes
\begin{equation}
\frac1{J}\sum_{k=1}^{K_1}\frac1{x_{1,k}}=-2\pi p\,.
\end{equation}
In particular, the rational three spin configuration corresponds to the solutions where
$
n_k=n$ for all the values $k\in \{1,2,\ldots K_1\} $ and
$m_k=n$ for all the values $k\in \{1,2,\ldots K_2\}
$.
This set of equations can be solved using the same technique as it has been done for the 
undeformed $\grSU(3)$ case in \cite{Freyhult:2005fn}. For details we refer the reader to Appendix \ref{subsec:Solving the Bethe equations}.
Finally, the energy is given by 
\be
\label{energy_bethe}
E &=& 
 \frac{\lambda}{2 J}\Big( 4 m^2 j_1j_2+(m-n)^2j_1j_3
+(m+n)^2j_2j_3 
\nn \\
&+& \bar{\sigma}^2 (j_2j_3+j_2j_1+j_3j_2-j_1j_2j_3) \Big) \,,
\ee
where the normalized spin $j_i$ is defined as $j_i=J_i/J$. Thus, the circular string energy (\ref{Energy:string}) obtained from the string $\grSU_q(3)$ $\sigma$-model in section \ref{subsec:circular_string_sol_v2} agrees with the energy \eqref{energy_bethe} computed from the Bethe equations \eqref{bethe1} and \eqref{bethe2}.  
In particular, this confirms the presence of the so-called ``cubic term'': The contribution $\bar{\sigma}^2  j_1 j_2 j_3$ in \eqref{energy_bethe}.


\section{Conclusions}
\label{sec:conclusions}

In this letter the fast spinning string $\sigma$-model corresponding to the mixed and deformed $\grSU_q(3)$ sector
of the dual gauge theory, has been derived for the case of purely imaginary $\beta$ deformations. 
For the special rational string solution, we have computed the energy from the $\grSU_q(3)$ $\sigma$-model, and compared with the solutions predicted by the deformed Bethe equations at strong coupling. We have found perfect agreement between the two expressions.
%
%
It is also clear from our approach that an extra cubic term is necessary in the coherent state Hamiltonian \eqref{coherent_H}. 

The initial $\grSU(3)$ symmetry is broken in the deformed theory. Indeed, the symmetries are now encoded in the $\algSU_q(3)$ quantum algebra. Therefore, instead of \eqref{n}, a different basis that respects $\algSU_q(3)$ should be chosen for the coherent states. Thus, in this perspective it is not surprising that the more naive derivation of the coherent state Hamiltonian misses a contribution at the quadratic order in the deformation parameter $\bar\sigma$. Notice that in \cite{Frolov:2005iq}, this problem was solved by performing a non-unitary transformation on the spin-chain Hamiltonian in order to obtain the correct $\sigma$-model. In principle, it should be possible to apply the same technique in our case. But finding the proper transformation is difficult in the general case.
In order to determine the continuum limit of the spin chain Hamiltonian, more knowledge about the corresponding supergravity background would be helpful. Such continuum regimes of discrete spin chain models play an important role not only in string theory, in the context of AdS/CFT duality, but also in numerous other contexts, \emph{e.g.} in condensed matter physics and in statistical mechanics. 
Coherent states have been constructed for $\grSU(2)_q$ \cite{Aizawa:2009} and also  $\grSU(3)_q$ \cite{Providencia:1993,Li:1994} quantum groups. However, they are not all equivalent and it is not clear 
how to obtain a continuum limit action from them.

In~\cite{Mansson:2008xv} it was found that the one-loop dilatation operator for the holomorphic Leigh-Strassler deformed $\grSU(3)$ spin chain with simultaneously non-zero deformation parameters $q$ and $h$, cf.~\eqref{superpotential}, has an $\algSU_{q,h}(3)$ Hopf symmetry algebra, which is not of a quasi-triangular type.
For this case, with non-zero $h$ and $q$, the string dual remains unknown, even though a perturbative method was developed in \cite{Kulaxizi:2006zc}, where the gravity dual
was constructed up to third order in the parameter $h$.
A general procedure for deriving continuum limits of discrete models with a given quantum group symmetry might help in order to make an educated guess to determine the string $\sigma$-model for Leigh-Strassler deformations with general parameters $q$ and $h$.

Another natural and interesting direction would be the construction of a Lax pair for the string $\grSU_q(3)$ $\sigma$-model analyzed in this letter. In this case we have a set of Bethe equations, and we expect that a Lax pair for the corresponding $\sigma$-model exists. This should give rise to the same Bethe equations in the thermodynamic limit at strong coupling. Finally, it would be interesting to investigate the mixed $\grSU_q(3)$ sector in the case of real $\beta$.

 \section*{Acknowledgements} 
 
We thank L. Freyhult, S.F. Hassan, T. Klose, M. Pawellek and A.~Tseytlin for valuable discussions. In particular we are thankful to K. Zoubos not only for the numerous and helpful discussions but also for a careful reading of the manuscript. We are also grateful to 
T. M. S$\o$rensen for reading parts of the manuscript.
The work of T. M\aa{}nsson was financed
by the Swedish Science Research Council.

\begin{appendix}

\section{Geometrical set-up}
\label{sec:geom_setup}

Here we collect some useful expressions about the Lunin-Maldacena geometry, mostly for completeness and for setting down the notation.
The Lunin-Maldacena metric for the deformed $\ads_5\times \sphere^5$ is
\be
\label{metric}
ds^2= R^2 \, H^{1/2} \left[ 
            ds^2_{\ads_5}
            +\sum_{i=1}^3 (d\rho_i^2+G\rho_i^2 d\phi_i^2)
            +(\tilde\gamma^2+\tilde\sigma^2)G \rho_1^2\rho_2^2\rho_3^2 (\sum_{i=1}^3 d\phi_i)^2\right]\,,\quad
 \ee           
where 
\be
&& H=1+\tilde\sigma^2 Q\,,
\quad
 G^{-1} \equiv 1+(\tilde\gamma^2+\tilde\sigma^2) Q\,,
 \quad
 Q \equiv \rho_1^2\rho_2^2+\rho_2^2\rho_3^2+\rho_1^2\rho_3^2\,.
\quad\ee
The deformation parameter is
\be
\tilde\beta\equiv \beta R^2 =\tilde\gamma -i \tilde\sigma\,.
\ee
The B-field is defined by
\be
B_2\,=\, - R^2\, \left( \tilde\gamma \, G\,  w_2- 12\, \tilde\sigma \, w_1 \,d\psi \right)\,,
\ee
with
\be
\label{eq:oneforms}
&& w_2 = \rho_1^2\rho_2^2 d\phi_1\wedge d\phi_2 + \rho_2^2\rho_3^2 d\phi_2\wedge d\phi_3 +\rho_1^2 \rho_3^2 d\phi_3 \wedge d\phi_1\,, 
\qquad
\psi = {1\over 3} \left ( \phi_1 +\phi_2+\phi_3\right)\,
\nn\\
&& w_1 ={1\over 4} \rho_1 \rho_3 \left( \rho_3 d\rho_1-\rho_1d\rho_3\right)\,.
\ee 
Notice that actually $w_1$ is defined only up to a constant, explicitly
\be
\label{def_w1}
d w_1 \equiv \cos \theta \sin^3 \theta \sin\psi \cos\psi \, d\theta \wedge d\psi\,.
\ee
Our choice for $w_1$ in \eqref{eq:oneforms} reduces to \eqref{def_w1} once one inserts the parameterization \eqref{parameterization}, which is the one adopted in the whole paper.
%

\paragraph{Our set-up.}
We are investigating the case when the deformation is purely imaginary, namely for us 
\be
\tilde\gamma=0\,, \qquad \tilde\beta=-i \tilde\sigma\,, \qquad q= e^{2 \pi i \beta}\,.
\ee

\paragraph{Parameters.}
We recall also the relations among the string tension and the 't Hooft coupling constant $\lambda$ due to the $\ads_5/\mbox{CFT}_4$ correspondence, as well as our convention in setting $\alpha'=1$, i.e.
\be
{R^2 \over \alpha'}= \sqrt{\lambda}\,, \qquad \alpha'=1
\,, \qquad
\tilde \lambda = {\lambda \over J^2}= {1\over \mathcal{J}^2}\,. 
\ee


\section{$\grSU(3)_\beta$ string $\sigma$-model}
\label{sec:explicit_derivation}

In this section we report a more detailed derivation of the $\grSU_q(3)$ string $\sigma$-model. In particular, in the appendix we follow the approach used in \cite{Kruczenski:2004cn, Frolov:2005ty, Frolov:2005iq} for the $\grSU_{\beta}(2)$ sector.
The classical string configuration we are describing is a string moving at the center of $\ads_5$ and in the $\grSU(3)_q$ sub-sector of $\sphere^5$.


\subsubsection*{The Lagrangian }
\label{sec:Lagrangian}

The bosonic string action is 
\be
\label{def_bosnicaction}
S_{B} &=& \sqrt{\lambda} \int d\tau \int_0^{2\pi} {d\sigma\over 2\pi} \CL_B\,= \int d\tau \int_0^{2\pi} {d\sigma\over 2\pi}\left( \CL_{\rm kin}+\CL_{\rm WZW}\right)
 \nn\\ &=&
-{1\over 2\alpha' } \int d\tau \int_0^{2\pi} {d\sigma \over 2\pi} 
\left [ \gamma^{\alpha\beta} \p_\alpha X^M \p_\beta X_N G_{MN}
        -\epsilon^{\alpha\beta} \p_\alpha X^M \p_\beta X^N B_{MN}
        \right] 
\ee
where $\gamma^{\alpha\beta}= \sqrt{-h}h^{\alpha\beta}$, Greek letters $\alpha,\, \beta=0,1$ label the world-sheet indices, Latin capital letters $M,N=0,\dots, 9$ label the curved target space indices. 
For a string moving on $t\in \ads_5$ and on $\grSU(3)\subset\grSU(4)$ the relevant terms in the initial metric \eqref{metric} are
\be
ds^2= R^2\, H^{\half} 
           \left[ -dt^2+ 
           +\sum_{i=1}^3 (d\phi_i^2+G\rho_i^2 d\phi_i^2)
            +(\tilde\gamma^2+\tilde\sigma^2)G \rho_1^2\rho_2^2\rho_3^2 (\sum_{i=1}^3 d\phi_i)^2\right]\,,
            \quad
 \ee       
%
%
The parameterization suitable to describe the mixed $\grSU_q(3)$ sector is in \eqref{parameterization}.
%
%
%
%
%
By using \eqref{parameterization} the Lagrangian in \eqref{def_bosnicaction} becomes
\be
\label{new_new_lkin}
\CL_{\rm kin}
&=& 
-{1\over 2}\gamma^{\alpha\beta} \p_\alpha X^M \p_\beta X_N G_{MN}
 \nn\\
&=&
-{1 \over 2} \sqrt{H} \gamma^{\alpha\beta} \big\{
           -\p_\alpha t\p_\beta t
           +\p_\alpha\th \p_\beta \th 
           +\sin^2 \th \p_\alpha \psi \p_\beta \psi
            \nn\\
           &+&G \big[ \p_\alpha\xi \p_\beta \xi
           + \sin^2 \th \p_\alpha \varphi_1 \p_\beta \varphi_1 
           + \cos^2 \th \p_\alpha\varphi_2 \p_\beta \varphi_2
           +2  \cos^2 \th \p_\alpha \xi \p_\beta \varphi_2 
           \nn \\
           &+& 2 \cos{2\psi} \sin^2 \th \p_\alpha \xi \p_\beta \varphi_1\big]
            \nn\\ 
           &+& G\, \tilde\sigma^2 \cos^2\th \sin^4\th \cos^2\psi \sin^2\psi \big[
           \p_\alpha\xi \p_\beta \xi
            +2 \p_\alpha \xi \p_\beta \varphi_2 
           +4 \p_\alpha\xi \p_\beta \varphi_1 
            \nn\\
           &+ &\p_\alpha \varphi_2 \p_\beta \varphi_2 
           +4 \p_\alpha \varphi_1 \p_\beta \varphi_1 
           + 4 \p_\alpha \varphi_1 \p_\beta \varphi_2 
           \big] 
           \big\}
\\
\label{new_new_lwzw}
\CL_{\rm WZW}
&=&
{1\over 2} \epsilon^{\alpha\beta} \p_\alpha X^M \p_\beta X^N B_{MN}
\nn\\
&= &- {1 \over 2} \tilde\sigma \sin{2\th} \epsilon^{\alpha\beta} \big[
\p_\alpha \xi + 2 \p_\alpha \varphi_1 +\p_\alpha \varphi_2\big]
\big[ \cos^2 \psi \p_\beta \th -\quarter \sin{2\th} \sin{2\psi} \p_\beta \psi \big]\qquad
\ee

\subsubsection*{T-dualizing}
\label{subsec:Tduality}

In order to T-dualize along the coordinate $\alpha$, one substitutes the term $\p_\alpha \xi$ with an auxiliary field $A_\alpha$ and adds the Lagrange multiplier term $- \epsilon^{\alpha\beta} A_\alpha \p_\beta \tilde\xi$. 
Then one needs to solve the equations of motion for $A_\alpha$, the solution in this case reads 
\be
\label{sol_A}
 A_\alpha &=&
 \p_\alpha \varphi_2  (-1 +D \sin^2 \th) 
- \p_\alpha \varphi_1 D \sin^2 \th (\cos{2\psi} + {1\over 16} \tilde\sigma^2\sin^2{2\th} \sin^2{2\psi}) 
\nn\\ &-&
 {h_{\alpha\beta}\over\sqrt{-h}} \epsilon^{\beta\gamma} {D\over G H \sqrt{\lambda}} \p_\gamma \tilde\xi 
 \nn \\
&-&\half {h_{\alpha\beta}\over\sqrt{-h}} \epsilon^{\beta\gamma} \tilde\sigma {D\over G \sqrt{H}} \sin{2\th} 
   ( \cos^2\psi \p_\gamma \th -\quarter \sin{2\psi} \sin{2\th} \p_\gamma \psi)\,,
\ee
where D is defined as
\be
\label{def_D}
D\equiv \frac{1}{1 +\tilde\sigma^2\sin^4\th \cos^2\th \cos^2\psi \sin^2\psi}\,.
\ee
By substituting back the expression for $A_\alpha$ \eqref{sol_A} into the Lagrangian, one obtains the final T-dual Lagrangian, which is the Lagrangian that we will use for imposing the static gauge and expand in the fast string coordinates. 
Thus, the T-dual Lagrangian is
\be
\label{Tdual_Lkin}
\tilde{\CL}_{\rm kin} &=&
-{1\over 2} \sqrt{H} \gamma^{\alpha\beta} \big\{
-\p_\alpha t \p_\beta  t
+\p_\alpha \th \p_\beta \th 
+\sin^2\th \p_\alpha \psi \p_\beta \psi
+G \big[ \cos^2\th \sin^2\th \p_\alpha\varphi_2 \p_\beta \varphi_2
\nn\\ 
          &-&  {D\over 4} \sin^2\th (-3 - \cos{2\th} +2 \cos{4\psi} \sin^2 \th ) \p_\alpha\varphi_1 \p_\beta \varphi_1
            \nn \\
            & - &2 D \cos^2\th \sin^2\th\cos{2\psi} \p_\alpha\varphi_1 \p_\beta \varphi_2\big]
 \nn \\              
&+&{\tilde\sigma^2 \over 2}G D \cos^2\th \sin^4\th \sin^2{2\psi} \big[
            (9-\cos{2\th}-8 \cos{2\psi} \sin^2\th) \p_\alpha\varphi_1 \p_\beta \varphi_1
            \nn \\
           & -& \sin^2\th (\cos{2\psi} -2)\p_\alpha\varphi_1 \p_\beta \varphi_2\big]
 \nn\\          
&+& {\tilde \sigma^2\over 4} {D\over G H} \sin^2{2\th} 
    (\cos^2\psi \p_\alpha \th -\half \sin{2\th} \sin\psi \cos\psi \p_\alpha \psi)   
    \nn \\         
     &\times& (\cos^2\psi \p_\beta \th -\half \sin{2\th} \sin\psi \cos\psi \p_\beta \psi)  
 \nn\\    
&+& {1\over\lambda} {D\over H G} \p_\alpha \tilde\xi  \p_\beta  \tilde\xi
- \tilde\sigma {D \over G H} \p_\alpha\tilde \xi (-\cos^2\psi \p_\beta \th+\quarter \sin{2\th} \sin{2\psi}\p_\beta \psi)\big\}
%
\\ 
\label{Tdual_LWZW}
\tilde{\CL}_{\rm WZW}& =&
{1\over 2}\tilde\sigma \sin{2\th}\,(-2 +\cos{2\psi}\sin^2\th) D \,
      \epsilon^{\alpha\beta}\p_\alpha\varphi_1\,(\cos^2\psi \p_\beta \th-\quarter \sin{2\th} \sin{2\psi}\p_\beta \psi)
 \nn\\ 
&-& {1\over 2}\tilde\sigma \,  \sin{2\th} \, \sin^2\th  D \,
      \epsilon^{\alpha\beta}\p_\alpha\varphi_2\, (\cos^2\psi \p_\beta \th-\quarter \sin{2\th} \sin{2\psi}\p_\beta \psi)
      \nn\\ 
&+& {D\over \sqrt{\lambda}}\,  \epsilon^{\alpha\beta}\p_\beta\tilde\xi \, \Big[
      \left( -\cos{2\psi} \sin^2\th - {\tilde\sigma^2\over 4} \cos^2\th \sin^4\th\left (-1+\cos{4\psi}\right) \right)\p_\alpha \varphi_1
       \nn\\
      &+& \cos^2\th \,\left(1+\tilde\sigma^2 \cos^2\psi\, \sin^4\th\, \sin^2\psi\right)\p_\alpha\varphi_2\Big]
\ee

\subsubsection*{Static gauge and fast string expansion}
\label{subsec:faststring}

We impose the static gauge according to \eqref{static_gauge_true}.
For the kinetic term it is convenient to use the Nambu-Goto action 
\be
\CL_{\rm kin}= -\half \sqrt{-h} h^{\alpha\beta} g_{\alpha\beta}= - \sqrt{-\det_{\alpha,\beta}g_{\alpha\beta}}\,,
\ee
where $g_{\alpha\beta}$ is the induced metric on the world-sheet, i.e.
$ 
g_{\alpha\beta}= \p_\alpha X^M \p_\beta X^N G_{MN}
$\,.

In order to consider the fast limit one needs to rescale the time derivatives through the effective parameter $\tilde \lambda = {\lambda \over J^2}$
\be
\label{rescaling_faststring}
\tau \rightarrow {1\over\tilde\lambda} \tau \,, \qquad \p_0 \rightarrow \tilde \lambda \p_0 
\,, \qquad
\tilde \sigma\rightarrow \sqrt{\tilde\lambda} \tilde \sigma\,,
\qquad
 {\tilde\beta \over\sqrt{\tilde{\lambda}}}  = \text{fixed} \,, \quad \text{for} \quad \tilde\lambda\rightarrow 0\,, \qquad
\ee
and then consider the small $\tilde\lambda$ limit, i.e. $\tilde\lambda\equiv{\lambda\over J^2}\equiv {1\over \mathcal J^2} \rightarrow 0$. Notice that it is necessary to keep the deformation parameters fixed under the fast string expansion (and it is in general true also for the real part of $\tilde\beta$, i.e. $\tilde\gamma$) in order to have a consistent theory. 
%
Hence, after imposing the gauge \eqref{static_gauge_true}, the rescaling \eqref{rescaling_faststring}, and taking the limit $\tilde\lambda\rightarrow 0$, the result for the Lagrangian at the leading order in $\tilde\lambda$ is
\be
\label{new_final_Lkin}
\tilde \CL_{\rm kin} &=& -\sqrt{\det_{\alpha\beta} g_{\alpha\beta}}=
-{ \lmtilde\over 2} \Big[ 
(\th')^2 
+\sin^2\th (\psi')^2
-\half \cos{2\psi} \sin^2{2\th} \varphi_1' \varphi'_2
\nn\\
&+&\quarter \sin^2{2\th} (\varphi_2')^2
+\sin^2\th (\cos^2\th +\sin^2{2\psi} \sin^2\th)  (\varphi_1')^2
\nn \\
&+& \bar\sigma \sin{2\th} \big( \cos^2\psi \th' -\quarter  \sin{2\th} \sin{2\psi} \psi' \big)
\nn\\ 
&+&\bar\sigma^2 \sin^2\th \big( \cos^2\th+\quarter  \sin^4\th  \sin^2{2\psi} \big)\Big] 
+\mathcal O(\lmtilde^2)
%
\\ 
\label{new_final_LWZW}
\tilde\CL_{WZW} &=&  \cos^2\th \dot \varphi_2 +\cos{2\psi} \sin^2\th \dot\varphi_1 +\mathcal O(\lmtilde^2)
\ee
\section{Solving the Bethe equations: details}
\label{subsec:Solving the Bethe equations}

In this appendix we report the details for solving the Bethe equations.
For reader's convenience, we rewrite the equations by using
\be
X  &:=&
 \frac{\sinh((\mu_{2,k}-\mu_{2,l})-2\pi\sigma)}{\sinh((\mu_{2,k}-\mu_{2,l})+2\pi\sigma)}
 \nn \\
& = &\frac{\tanh \mu_{2,k}-\tanh \mu_{2,l}-\tanh 2\pi\sigma
(1-\tanh \mu_{2,k}\tanh \mu_{2,l})}
{\tanh \mu_{2,k}-\tanh \mu_{2,l}+\tanh 2\pi\sigma
(1-\tanh \mu_{2,k}\tanh \mu_{2,l})} 
\nn \\
&=&\frac{x_{2,k}-x_{2,l}+\frac{i \tanh 2\pi\sigma}{J \tanh \pi\sigma}
(1+4\tanh^2 (\pi\sigma)J^2x_{2,k}x_{2,l})}
{x_{2,k}-x_{2,l}-\frac{i \tanh 2\pi\sigma}{J \tanh \pi\sigma}
(1+4\tanh^2 (\pi\sigma )J^2x_{2,k}x_{2,l})}\,,
\ee
where $x_{2,k}$ is defined through
 $\tanh \mu_{2,k}=2i \tanh (\pi\sigma) Jx_{2,k}$. Expanding the
logarithm of $X$ in 
small $\sigma$ and large $J$ and setting $\bar{\sigma}=\sigma J$ finite, we obtain
\begin{equation}
\log X=\frac{i2}{J}\frac{1+(2\pi \sigma J)^2x_{2,k}x_{2,l}}{x_{2,k}-x_{2,l}}\,.
\end{equation}
Now we take the logarithm and repeat the same expansion with the equations
\eqref{bethe1}, \eqref{bethe2} and \eqref{bethe_energy1}. Thus, in this limit the energy reads
\be
\label{Spin-energy}
E= \frac{\lambda}{8\pi^2 J^2} \left(\sum_k^{K_1}\frac1{x_{1,k}^2}+(2\pi\bar{\sigma} )^2K_1 \right)\,.
\ee
 The equations \eqref{bethe1} and \eqref{bethe2} become
\begin{equation}
\label{eq:1}
2\pi m_k=\frac{2}{J} \sum_{l\neq k}^{K_2}
\frac{1+(2\pi\bar{\sigma} )^2x_{2,k}x_{2,l}}{x_{2,k}-x_{2,l}}
-\frac{1}{J}\sum_{j=1}^{K_1} 
\frac{1+(2\pi\bar{\sigma} )^2x_{2,k}x_{1,j}}{x_{2,k}-x_{1,j}} \,,
\end{equation}
\bea
\label{eq:2}
 &&\frac{1}{x_{1,m}}+2\pi n_m=
\frac{2}{J}\sum_{l\neq m}^{K_1} 
\frac{1+(2\pi\bar{\sigma} )^2x_{1,m}x_{1,l}}{x_{1,m}-x_{1,l}}
\,
-\frac{1}{J}\sum_{j=1}^{K_2} 
\frac{1+(2\pi\bar{\sigma} )^2x_{2,k}x_{1,j}}{x_{1,m}-x_{2,j}}\,, 
\eea
where $n_k$ and $m_k$ are integers.
The rational three spin solution corresponds to
\be
\begin{split}
&n_k=\tilde{n} \quad \forall \quad k\in \{1,2,\ldots K_1\} \\
& m_k=\tilde{m} \quad\forall \quad k\in \{1,2,\ldots K_2\} \,.
\end{split}
\ee
In this expansion the cyclicity constraint becomes
\begin{equation}
\frac1{J}\sum_{k=1}^{K_1}\frac1{x_{1,k}}=-2\pi p\,.
\end{equation}
This set of equations can be solved in the same fashion as was done for the 
$\grSU(3)$ case in \cite{Freyhult:2005fn}.
We can immediately obtain a relation for $p$ if we
sum the first equation \eqref{eq:1} over $k$ running from $1$ to $K_2$ and sum the second
equation over $k$ running from $1$ to $K_1$  \eqref{eq:2} and add them both, namely
\begin{equation}
p=\tilde{m}\frac{K_2}{J}+\tilde{n}\frac{K_1}{J}\,.
\end{equation}
By multiplying the first equation \eqref{eq:1} and the second \eqref{eq:2} by 
\begin{equation}
\frac1{J^2}\sum_{k=1}^{K_2}\sum_{m=1}^{K_1}
\frac{1+(2\pi\bar{\sigma} )^2x_{2,k}x_{1,m}}{x_{2,k}-x_{1,m}}\,,
\end{equation}
and then adding the result, we end up with the following equation
\begin{equation}
\label{eq:ABCD}
C+2\pi(\tilde{m}+\tilde{n})D=A+B\,,
\end{equation}
where $A$, $B$, $C$ and $D$ are defined as
\be
A &=&-\frac1{J^3}\sum_{k=1}^{K_2}\sum_{m=1}^{K_1}\sum_{l=1}^{K_1}
\frac{(1+(2\pi\bar{\sigma} )^2 x_{2,k}x_{1,m})(1+(2\pi\bar{\sigma} )^2 x_{2,k}x_{1,l})}
{(x_{2,k}-x_{1,m})(x_{2,k}-x_{1,l})} 
\nn \\
&-&\frac{2}{J^3}\sum_{k=1}^{K_2}\sum_{m=1}^{K_1}\sum_{l\neq m}^{K_1}
\frac{(1+(2\pi\bar{\sigma} )^2 x_{2,k}x_{1,m})(1+(2\pi\bar{\sigma} )^2 x_{2,k}x_{1,l})}
{(x_{2,k}-x_{1,m})(x_{1,m}-x_{1,l})}=
\nn \\
&=&- \frac1{J^2}\sum_{k=1}^{K_2}\sum_{m=1}^{K_1}
\frac{1+2(2\pi\bar{\sigma} )^2x_{1,m}x_{2,k}}{(x_{2,k}-x_{1,m})^2}
+\frac{(2\pi\bar{\sigma} )^2}{J^3}\sum_{m=1}^{K_2}\sum_{k=1}^{K_1}\sum_{l\neq k}^{K_1}1\,,
\ee
\be
B &=&\frac1{J^3}\sum_{m=1}^{K_1}\sum_{k=1}^{K_2}\sum_{l=1}^{K_2}
\frac{(1+(2\pi \bar{\sigma} )^2 x_{1,m}x_{2,k})(1+(\sigma L)^2 x_{1,m}x_{2,l})}
{(x_{2,k}-x_{1,m})(x_{1,m}-x_{2,l})}\nn \\
&+& \frac{2}{J^3}\sum_{m=1}^{K_1}\sum_{k=1}^{K_2}\sum_{l\neq k}^{K_2}
\frac{(1+(2\pi \bar{\sigma} )^2 x_{2,k}x_{1,m})(1+(2\pi\bar{\sigma} )^2 x_{2,k}x_{2,l})}
{(x_{2,k}-x_{1,m})(x_{2,k}-x_{2,l})}=\nn \\
&= &\frac1{J^2}\sum_{m=1}^{K_2}\sum_{k=1}^{K_1}
\frac{1+2(2\pi\bar{\sigma})^2x_{2,m}x_{2,k})^2}{(x_{2,m}-x_{1,m})^2}
-\frac{(2\pi\bar{\sigma})^2}{J^3}\sum_{m=1}^{K_1}\sum_{k=1}^{K_2}\sum_{l\neq k}^{K_2}1\,,
\ee
\begin{equation}
\begin{split}
C=&\frac1{J^2}\sum_{m=1}^{K_2}\sum_{k=1}^{K_1}\frac1{x_{1,k}}
\frac{1+(2\pi\bar{\sigma} )^2x_{1,k}x_{2,m}}{x_{2,m}-x_{1,k}}\,,
\end{split}
\end{equation}
\begin{equation}
D=\frac1{J^2}\sum_{m=1}^{K_2}\sum_{k=1}^{K_1}
\frac{1+(\sigma L)^2x_{1,k}x_{2,m}}{x_{2,m}-x_{1,k}}\,.
\end{equation}
We are considering the limit when there is a large number of roots such that the
$K_i$ goes as $J$, which is large, then at the leading order the sums are 
\begin{equation}
\frac1{J^3}\sum_{m=1}^{K_2}\sum_{k=1}^{K_1}\sum_{l\neq k}^{K_1}1=
\frac{K_2K_1^2}{J^3}+\mathcal{O}(1/J) \,,\qquad
\frac1{J^3}\sum_{m=1}^{K_1}\sum_{k=1}^{K_2}\sum_{l\neq k}^{K_2}1=
\frac{K_2^2K_1}{J^3}+\mathcal{O}(1/J)\,.
\end{equation}
Using this we obtain
\be
\label{AB}
A+B=\frac{(2\pi\bar{\sigma})^2K_2K_1}{J^3}(K_1-K_2)\,,
\ee
\begin{equation}
\label{C}
\begin{split}
C=&\frac1{J}\sum_{k=1}^{K_1} \frac1{x_{1,k}^2}+4\pi^2 p^2
-4\pi^2 p \tilde{n}+(2\pi\bar{\sigma} )^2 \frac{K_1^2}{J^2} + \mathcal{O}(1/J)\,,
\end{split}
\end{equation}
\begin{equation}
\label{D}
D=-2\pi \tilde{m} \frac{K_2}{J}\,.
\end{equation}
Note that the sum in the expression for the energy \eqref{Spin-energy} is in the expression for $C$, thus we can use
the equation \eqref{eq:ABCD}.
Insert the expressions \eqref{AB}, \eqref{C}, and \eqref{D} into the equation (\ref{eq:ABCD}), we find for the energy
\be
\label{Energy-final}
E &=& \frac{\lambda}{8\pi^2} \frac1{J^2}\sum_{k=1}^{K_1}\frac1{x_{k,1}}+(2\pi\bar{\sigma} )^2 \frac{K_1}{J}=\nn\\
&=&\frac{\lambda}{8\pi^2 J}\left(\tilde{n}^2\frac{K_1}{J}(1-\frac{K_1}{J})+2\tilde{m}\tilde{n}\frac{K_2}{J}(1-\frac{K_1}{J})
+\tilde{m}^2\frac{K_2}{J}(1-\frac{K_2}{J}) \right. +\nn\\ 
&+& 
\left. \bar{\sigma}^2\left(\frac{K_2K_1}{J^3}(K_1-K_2)-\frac{K_1^2}{J^2}+ 
\frac{K_1}{J} \right)\right)\,.
\ee
Along the lines of \cite{Freyhult:2005fn}, we set $J_1=J-K_1$, $J_2=K_1-K_2$
and $J_3=K_2$. 
In order to facilitate the comparison with the string result, we redefine the constants according to
\be
\tilde{n}=2 n\,, \qquad \tilde{m}=-n-m\,,
\ee
and the energy can be written as
\be
E=\frac{\lambda}{8\pi^2 J}\left( m^2 j_1j_2+(m-n)^2j_1j_3
+(m+n)^2j_2j_3 +\bar{\sigma}^2 (j_2j_3+j_2j_1+j_3j_2-j_1j_2j_3)\right)\,.
\ee
Here we used the normalized spin $j_i=J_i/J$.

%
\section{Coherent $\sigma$ model}
\label{sec:coherent_sigma_mod}
Here we will show that using $\grSU(3)$ coherent states one obtains a continuum model which reproduce all quadratic terms, but 
the cubic term ($\rho_1^2\rho_2^2\rho_3^2$), cf \eqref{Hamiltonian:mixed}.%
\footnote{This result can be expected because we are using coherent states which do not respect the $\grSU_q(3)$ quantum symmetry. }

%

In the mixed $\grSU (3)$ sector a coherent state can be parameterized in the following way
\be
\label{n}
|n\rangle = \sin \theta \cos \psi e^{i\phi_1}|1\rangle +\sin \theta \sin \psi e^{-i\phi_2} |\bar{2}\rangle
+\cos \theta e^{i\phi_3}|3\rangle\,,
\ee
where $0\le \theta <\pi$, $0\le \psi < 2\pi$ and $0\le \phi_i < 2\pi$ for $i=1,2,3$. Note the opposite sign on the phase $\phi_2$.
%
%
The spin-chain Hamiltonian which represents the one-loop dilatation operator in the mixed deformed $\grSU_q(3)$ sector, has been shown to be
 integrable in \cite{Mansson:2007sh}. Apart from a phase, the spin chain Hamiltonian obtained in~\cite{Mansson:2007sh} differs from the
usual $\algSU_q(3)$ model, often called the trigonometric 
(or hyperbolic) $A_{2}$ vertex model~\cite{DeVega:1988ry}, by an additional term, which cancels out  
for periodic spin chains. 
In particular for $q=e^{2\pi \sigma}$ the Hamiltonian reads~\cite{Mansson:2007sh}
\be
\label{spinchain_h}
h_{i,i+1} &= &
\frac{2}{e^{-2\pi \sigma}+e^{2\pi \sigma}}\Big( \frac{e^{-2\pi \sigma}+e^{2\pi\sigma}}{2}(
E_{11}\otimes E_{\bar{2}\bar{2}}+E_{\bar{2}\bar{2}}\otimes E_{11}+E_{33}\otimes E_{\bar{2}\bar{2}}+E_{\bar{2}\bar{2}}\otimes E_{33})
\nn \\
&+&  e^{-2\pi\sigma}E_{11}\otimes E_{33}+e^{2\pi \sigma}E_{33}\otimes E_{11}
- ((E_{13}\otimes E_{31}+E_{1\bar{2}}\otimes E_{\bar{2}1}+E_{\bar{2}3}\otimes E_{3\bar{2}})
\nn\\ 
&+&h.c.\,)\Big)\,.
\ee
The operators $E_{ij}$ are defined through their action on the eigenstates $|i\rangle$,
$
E_{ij}|k\rangle = \delta_{jk} |i\rangle$. 
Since we are working at one-loop order, the spin-chain Hamiltonian only acts on the nearest neighboring sites.  
Starting from \eqref{spinchain_h} and using \eqref{n}, one can compute the matrix element
\be
\label{matrix_elem}
\mathcal H = {\lambda\over 8 \pi^2} \langle n_i\,, n_{i+1} | h_{i, i+1}| n_i\,, n_{i+1} \rangle\,+\mathcal{O}(\lambda)\,.
\ee
By assuming that the state does not change drastically from one site to the nearest one, {\it i.e.} $|n_{i+1}\rangle\cong |n_i +\delta_\sigma n_i\rangle$, up to 
quadratic terms in $\sigma$, the matrix element \eqref{matrix_elem} is 
\be
\label{coherent_H}
 \mathcal H 
&= & {\lambda \over 2 J}\Big\{
{\theta'}^2 
+\sin^2\theta (\cos^2\theta+\sin^2\theta \sin^2{2\psi}) {\varphi'_1}^2
- \half \sin^2{2\theta} \cos{2\psi} \varphi'_1 \varphi'_2
\nn \\
&+&\sin^2\theta \cos^2\theta {\varphi'_2}^2
+ \sin^2\theta\, {\psi'}^2
+ \bar{\sigma} \sin{2\theta} \left( \cos^2\psi\, \theta'-\quarter \sin2\theta \,\sin2\psi \,\psi' \right) 
\nn \\
&+ &\bar{\sigma}^2 \sin^2\theta \left( \cos^2\theta +\quarter \sin^2\theta \sin^22\psi \right)\Big\}+\mathcal{O}(\lambda),
\ee
where we define the finite $\bar{\sigma}=J \sigma$ with $J$ the length of the spin chain. The coherent state Hamiltonian
 \eqref{coherent_H} computed in this section is the Hamiltonian obtained from string computation \eqref{Hamiltonian:mixed} apart 
the cubic term $-\bar\sigma^2 \rho_1^2 \rho_2^2 \rho_3^3 $, as anticipated. We have checked that this Hamiltonian \eqref{coherent_H}
agrees with the Hamiltonian in \cite{Avan:2010xd}, which was obtained using a different method.

Note we had a freedom to add terms to the Hamiltonian \eqref{spinchain_h} which cancel out for a closed spin chain, these 
additional terms correspond to
different gauge choices of the B-field on the string theory side.

\end{appendix}

\bibliographystyle{newutphys}
\bibliography{leighref}








\end{document}